\documentclass[aps,pre,groupedaddress,showpacs,twocolumn]{revtex4-1}
\usepackage{graphicx,amsmath}
\usepackage{float}
\usepackage[absolute]{textpos}

\begin{document}

\title{Early warning signal for interior crises in excitable systems}

\author{Rajat Karnatak}
\email[]{karnatak@igb-berlin.de}
\affiliation{Leibniz-Institute of Freshwater Ecology and Inland Fisheries, 
M\"uggelseedamm 310, 12587, Berlin, Germany}

\author{Holger Kantz}
\email[]{kantz@pks.mpg.de}
\affiliation{Max Planck Institute for the Physics of Complex Systems, 
N\"othnitzer Stra\ss e 38, 01187 Dresden, Germany}

\author{Stephan Bialonski}
\email[]{bialonski@gmx.net}
\affiliation{Max Planck Institute for the Physics of Complex Systems, 
N\"othnitzer Stra\ss e 38, 01187 Dresden, Germany}

\begin{abstract}
The ability to reliably predict critical transitions in dynamical systems is a 
long-standing goal of diverse scientific communities. Previous work focused on 
early warning signals related to local bifurcations (critical slowing down) and 
non-bifurcation type transitions. We extend this toolbox and report on a 
characteristic scaling behavior (critical attractor growth) which is indicative 
of an impending global bifurcation, an interior crisis in excitable systems. We 
demonstrate our early warning signal in a conceptual climate model as well as in 
a model of coupled neurons known to exhibit extreme events. We observed critical 
attractor growth prior to interior crises of chaotic as well as 
strange-nonchaotic attractors. These observations promise to extend the classes 
of transitions that can be predicted via early warning signals.
\end{abstract}


\maketitle

\begin{textblock*}{20cm}(3cm,27cm)
 Published as Phys. Rev. E \textbf{96}, 042211 (2017). Copyright 2017 by the American Physical Society.
\end{textblock*}

The dynamical behavior of many complex systems can undergo dramatic changes. 
Such critical transitions have been associated with a multitude of complex 
systems that are important to human societies, ranging from potentially 
devastating transitions in the earth's climate system due to climate 
change~\cite{Lenton2011}, regime shifts such as desertification or 
eutrophication phenomena in 
ecosystems~\cite{Scheffer2015,Boettiger2013,Lenton2013}, to the progression 
into 
chronic disease states in humans (such as depression, inflammation, or 
cardiovascular disease)~\cite{Trefois2015}. Reliable and early predictions of 
critical transitions are very desirable and would not only allow for a warning 
of an impending event but also for time to prepare and to develop 
countermeasures.

Research that focusses on early warning signals for critical transitions 
associated with bifurcations (\emph{B-tipping}~\cite{Ashwin2012}) has gained 
strong momentum during the past 
years~\cite{Scheffer2012,Kuehn2011}, notably studies concentrating on the 
phenomenon of \emph{``critical slowing down"} (CSD)~\cite{Scheffer2009}. CSD 
manifests in slow rates of recovery from small perturbations near \emph{local} 
bifurcations, resulting in an increased variance and lag-1 autocorrelation in 
the time series of appropriate observables. 
CSD can be traced back to the fact that in local bifurcations, the dominant 
eigenvalue(s) of the Jacobians associated with equilibria or cycles smoothly 
approach(es) zero or the unit circle (leading to neutral/marginal stability), 
thereby slowing down and ultimately preventing the decay of perturbations at the 
bifurcation point.
While CSD has been demonstrated experimentally for different 
systems, including yeast populations~\cite{Dai2012} and food webs in 
lakes~\cite{Carpenter2011}, it has also been repeatedly pointed out to fail as 
an early warning signal if transitions are not associated with local 
bifurcations~\cite{Ditlevsen2010,Scheffer2012,Boettiger2013,Dakos2014}. 
Indeed, from a \emph{B-tipping} perspective, global bifurcations such as 
crises~\cite{Grebogi1983,Ditto1989}, which involve larger and more complex 
invariant sets compared to local bifurcations, are usually not tractable using 
local stability properties of equilibria and/or cycles. Thus, they are typically not 
revealed by CSD.

In this contribution, we demonstrate an early warning signal that is sensitive to a global bifurcation, namely an 
interior crisis in excitable systems. We consider crises in a conceptual 
climate 
model~\cite{Roberts2015} as well as in a system of coupled FitzHugh--Nagumo 
units~\cite{Ansmann2013,Karnatak2014b}. 
In both systems, we observe a characteristic scaling behavior to indicate an 
impending crisis, i.e. a sudden increase in 
the size of an attractor. In a second step, we exploit this scaling behavior in 
order to predict a crisis and, as a consequence, the onset of 
\emph{crisis-induced intermittency}~\cite{Grebogi1987} (post-crisis).
In excitable systems, the latter has been associated with dynamical regimes that 
can exhibit extreme events~\cite{Ansmann2013,Karnatak2014b,Ansmann2016}.

\begin{figure}
\includegraphics[width=8.6cm]{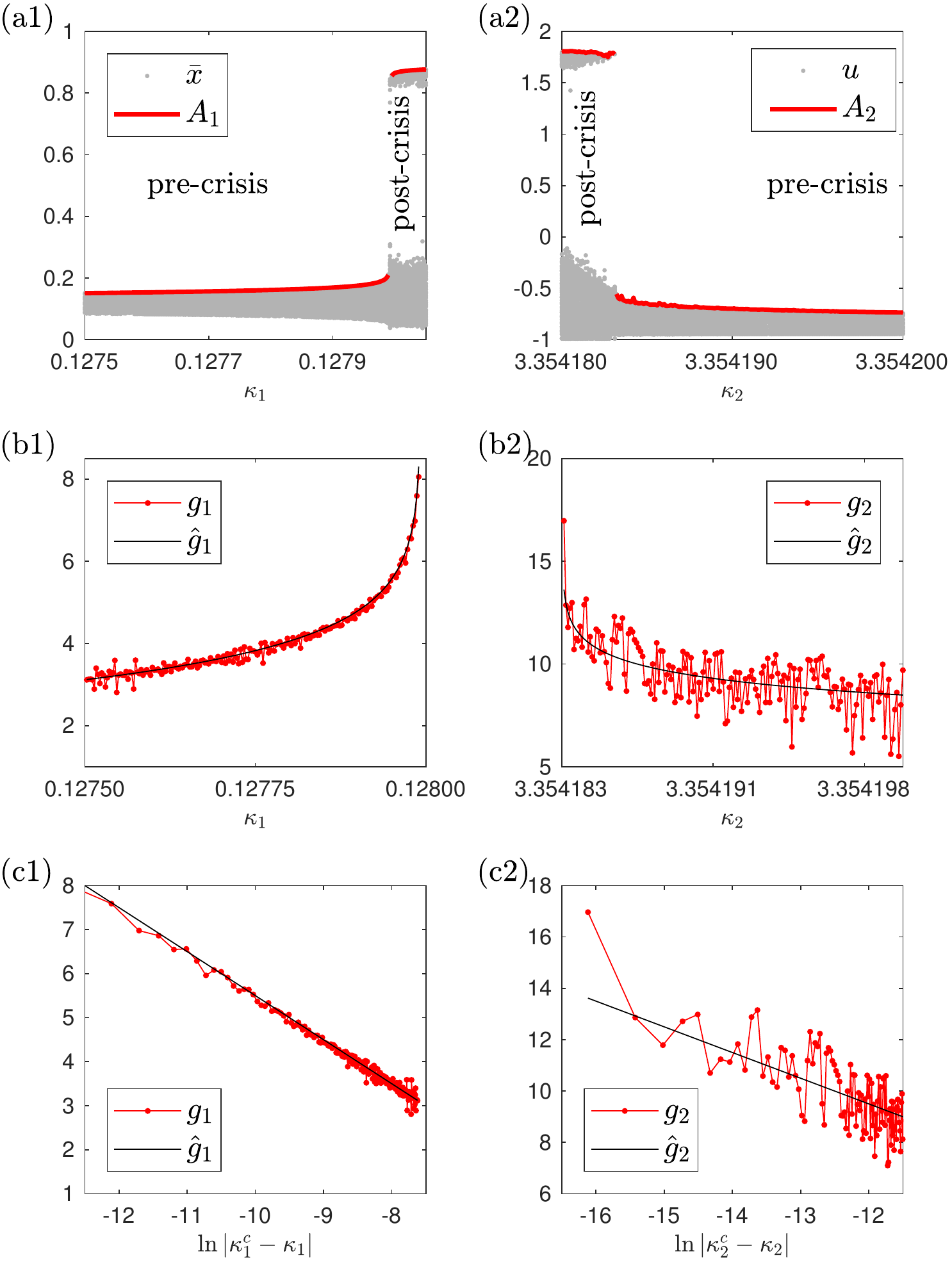}%
\caption{Bifurcation diagrams of System 1 (a1) and System 2 (a2). For each value 
of $\kappa_1$ and $\kappa_2$, grey points mark local maxima of a time series 
($\bar{x}(t)$ and $u(t)$, respectively) and red (dark) points denote the respective 
global maximum values, $A_1(\kappa_1)$ and $A_2(\kappa_2)$, determined from 
ensembles of time series 
(20 initial conditions, $10^5$ time units per time series for System~1; $10^3$ 
initial conditions, $10^7$ time units per time series for System~2; initial 
$10^5$ (System~1) and $10^6$ (System~2) time units were discarded as 
transients). 
Pre- and post-crisis regimes are marked by text labels. 
Parameter regimes of $\kappa_1$ $\in [0.1275, 0.12805]$ and $\kappa_2$ $\in 
[3.35418, 3.3542]$ were sampled in step sizes of $\delta\kappa_1=2.75 \times 
10^{-6}$ and $\delta\kappa_2=-1.0 \times 10^{-7}$, respectively. The set of 
equations~\eqref{eq:System1} and~\eqref{eq:System2} were solved 
numerically using the VODE package~\cite{BrownVODE1989}. Panels (b1) and (b2) 
show logarithmic growth rates (red-dotted curves) $g_1$ and $g_2$ as defined in 
equation~\eqref{eq:logampl} in the pre-crisis regimes of Systems~1 and~2 (i.e. 
$\kappa_1\in[0.1275,\kappa_1^c-\delta\kappa_1]$, $\kappa_2\in 
[\kappa_2^c+\delta\kappa_2,3.3542]$). The same data is shown as functions of the 
logarithm of the parametric distance from the critical values, $\ln{|\kappa_i^c 
- \kappa_i|}$, in panels (c1) and (c2) respectively. Lines $\hat{g}_1$ and 
$\hat{g}_2$ (black lines in panels (c1) and (c2)) were fitted to the data with a 
fixed slope of $-1$. These fits are also shown as black curves in panels (b1) 
and (b2).
\label{fig:1}}
\end{figure}

We consider the following two excitable systems to demonstrate the 
predictability of an interior crisis:
\begin{itemize}
\item System 1: Two mutually coupled FitzHugh--Nagumo oscillators as studied in 
Refs.~\cite{Ansmann2013,Karnatak2014b},
\begin{equation}
\begin{split}
\dot{x}_i & =  x_i (a-x_i) (x_i-1) - y_i + \kappa_1 (x_j - x_i), \\
\dot{y}_i & =  b_i x_i - c y_i,
\end{split}
\label{eq:System1}
\end{equation}
where $x_i$ and $y_i$ correspond to the excitatory and inhibitory variables 
respectively, and $i,j \in\{1,2\}, i \neq j$. Furthermore, let $\bar{x}(t)$ 
denote the arithmetic mean of the excitatory variables, 
$\bar{x}(t)=(x_1(t)+x_2(t))/2$. The coupling strength $\kappa_1$ was varied in 
our numerical studies, whereas the remaining parameters were fixed 
($a=-0.025794$, $c=0.02$, $b_1=0.0135$, $b_2=0.0065$).
\item System 2: Dimensionless form of a conceptual climate 
model~\cite{Roberts2015},
\begin{equation}
\begin{split}
\dot{u} &= \epsilon^{-1}\left(v-u^3+3u-k \right), \\
\dot{v} &=  p{(u-a)}^2-\kappa_2-mv-(\lambda+v)+w,\\
\dot{w} &=  r(\lambda+v-w),
\end{split}
\label{eq:System2}
\end{equation}
where the variable $u$ is related to the continental ice volume offset from a 
mean value, $v$ relates to the amount of carbon in the atmosphere, and $w$ to 
the amount of carbon in the mixed layer of the ocean. We varied parameter 
$\kappa_2$, while the remaining parameters were fixed at suitable values 
($\epsilon=0.1$, $a=0.8$, $p=3$, 
$k=4$, $r=0.05$, $m=1$, $\lambda=1$; see Ref.~\cite{Roberts2015}).
\end{itemize}

To demonstrate the characteristic scaling prior to crises in these systems, we 
show bifurcation diagrams of System~1 and System~2 in panels (a1) and (a2) of 
figure~\ref{fig:1}. 
In both diagrams, for each fixed parameter values of $\kappa_1$ and $\kappa_2$, 
red (dark) points represent maximum values $A_1$ and $A_2$ of $\bar{x}$ and $u(t)$, 
respectively, determined from ensembles of time series of different initial 
conditions. 
In System~1 (panel (a1)), the size of a chaotic attractor resulting from a 
period doubling route to chaos increases with increasing $\kappa_1$ values; the 
attractor undergoes an interior crisis. In System~2 (panel (a2)), a strange 
nonchaotic attractor which mediates a quasiperiodic route to chaos experiences 
an interior crisis for decreasing $\kappa_2$ values (similar to ``route A'' as 
described in Ref.~\cite{Witt1997}). In both cases, we observe $A_1$ and $A_2$ to 
show 
a steady growth as we approach the interior crisis, before exhibiting a 
discontinuous transition to high values beyond the critical values 
$\kappa_1^c\approx 0.1279922$ and $\kappa_2^c\approx 3.3541833$. 

To quantify the growth of the attractors for both systems in the pre-crisis 
regimes, we determine the logarithmic growth rate, 
\begin{equation}
g_i(\kappa_i)=\ln\left(\frac{A_i(\kappa_i+\delta\kappa_i)-A_i(\kappa_i)}{
\left|\delta\kappa_i\right|}\right), 
\label{eq:logampl} 
\end{equation} 
where $i\in\{1,2\}$ and $\delta\kappa_i$ are small increments of $\kappa_i$ (see 
caption of figure~\ref{fig:1}). The numerically determined growth rates $g_1$ 
and $g_2$ (panels (b1) and (b2) of fig.~\ref{fig:1}; red-dotted lines) of both 
systems for the pre-crisis regimes show a consistent increase with increasing 
proximity to the critical values. This proximity-dependent behavior becomes even 
more apparent if these growth rates are shown as functions of the logarithm of 
the parametric distance from the critical values, $\ln{|\kappa_i^c - 
\kappa_i|}$, in panels (c1) and (c2) {[}red dotted lines{]}. Indeed, these 
results suggest a linear scaling (with a slope of approximately $-1$ for our 
systems) between $g_i$ and the logarithmic parametric distances,
\begin{equation}
g_i(\kappa_i) \propto \ln\left|\kappa_i^c - \kappa_i\right|.
\end{equation}
We fitted lines $\hat{g}_1$ and $\hat{g}_2$ with a slope of $-1$ to the data in 
panels (c1) and (c2). These fits are shown as black curves in panels (b1)-(c2) 
and agree well with the observed data.

\begin{figure}[H]
\includegraphics[width=0.99\linewidth]{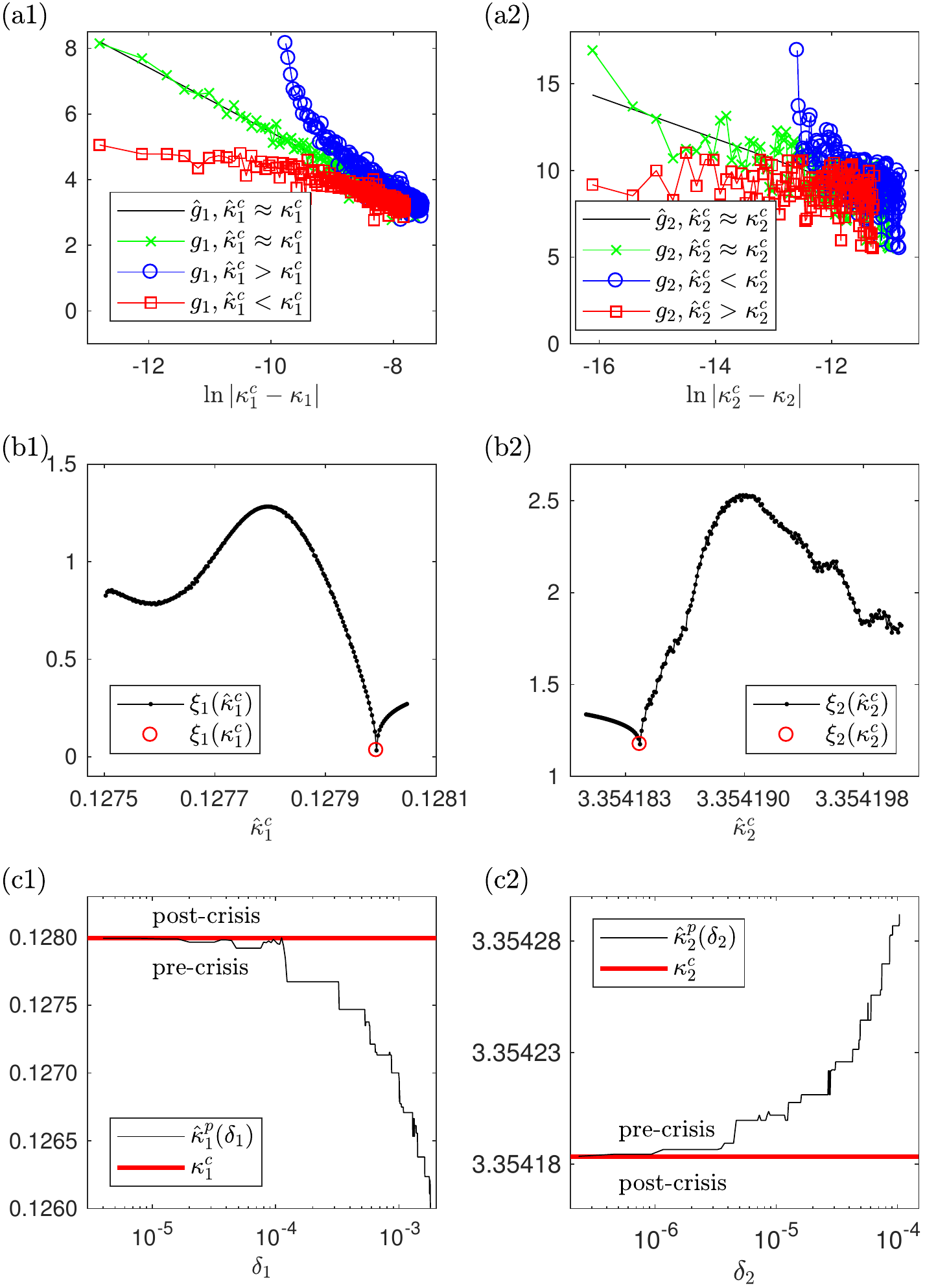}%
\caption{Logarithmic growth rates $g_1$ (a1) and $g_2$ (a2) as determined by 
equation~\eqref{eq:logampl} for values of $\kappa_1$ and $\kappa_2$ from the 
pre-crisis regimes of Systems~1 and 2, respectively. The growth rates are shown 
as functions of the logarithmic parametric distances to the candidate critical 
values $\hat{\kappa}_i^c$ of the systems. To determine growth rates, the 
ensembles of time series of figure~\ref{fig:1} were used. Different curves of 
$g_i$ are shown color-and-symbol coded and were obtained by varying the candidate critical 
values (red curves: $\hat{\kappa}_1^c = 0.127909$, $\hat{\kappa}_2^c = 3.35418$; 
blue curves: $\hat{\kappa}_1^c = 0.128047$, $\hat{\kappa}_2^c = 3.3541873$). 
Black lines denote linear fits to the curves for $\hat{\kappa}_i^c \approx 
\kappa_i^c$. The slopes of the best fits yielded $-0.983$ and $-1.190$, 
respectively, which is in approximate agreement with $-1$ as empirically 
observed in figure~\ref{fig:1}. 
(b1), (b2) Dependence of the mean squared errors $\xi_i$ of the linear fits on 
the candidate critical values $\hat{\kappa}_i^c$. Smallest errors were observed 
for those candidate critical values that equaled the actual critical values 
(marked by red circles). 
(c1), (c2) Dependence of the predicted critical values $\hat{\kappa}_i^p$ (narrow black 
lines) on the distance $\delta_i$ of the set of observed parameters with respect 
to the actual critical values $\kappa_i^c$ (thick red lines). 
For each value of $\delta_i$, the sets $J_i$ of observed parameters were created 
as $J_{i} = {[}\nu_i, \nu_i + \delta\tilde{\kappa}_i, \ldots, \nu_i + 
49\delta\tilde{\kappa}_i {]}$, respectively, where $\nu_{1/2} = \kappa_{1/2}^c 
\mp \delta_{1/2}$, $\delta\tilde{\kappa}_1=-4\cdot 10^{-6}$, and 
$\delta\tilde{\kappa}_2=2.34\cdot10^{-7}$.
\label{fig:2}}
\end{figure}

Can we predict an interior crisis prior to its occurrence by capitalizing on the 
scaling behavior? As stated before, crisis-induced 
intermittency~\cite{Grebogi1987} has been recently interpreted with respect to 
dynamical regimes that allow for the occurrence of extreme 
events~\cite{Ansmann2013,Karnatak2014b}. In this context, predicting an interior 
crisis would mean to predict the transition into a dynamical behavior which 
exhibits extreme events. To study whether an interior crisis can be predicted, 
we assumed that we do not know the actual values of $\kappa_1^c$ and 
$\kappa_2^c$; and we considered only data that was observed in the pre-crisis 
regimes (cf. figure~\ref{fig:1}). Let $\hat\kappa_1^c$ and $\hat\kappa_2^c$ 
denote our guesses of the critical values (from now on called ``candidate 
critical values''). In panels (a1) and (a2) of figure~\ref{fig:2}, we show the 
growth rates $g_i$ (as defined by equation~\eqref{eq:logampl}) as functions of 
the logarithmic parametric distance from the candidate critical values 
$\hat\kappa_i^c$ for System~1 and 2, respectively. For different candidate 
critical values of $\hat\kappa_i^c$, we obtained different curves. When the 
candidate critical values were located in the actual pre-crisis regimes 
($\hat\kappa_1^c < \kappa_1^c$, $\hat\kappa_2^c > \kappa_2^c$), the growth rates 
$g_i$ did not show a linear scaling relationship with respect to the logarithmic 
parametric distance to the candidate critical values. Instead, we observed 
concave downward curves (red curves with squares in panels (a1) and (a2)). As 
$\hat\kappa_i^c$ approached the actual critical values, we observed a transition 
from concave downward curves to a linear scaling behavior (with a slope of 
approximately $-1$; green crossed curves in panels (a1) and (a2)) for $\hat\kappa_i^c$ 
arbitrarily close or equal to the actual critical values. In addition, we show 
linear fits $\hat{g}_i$ (black lines) in both panels that emphasize the linear 
scaling behavior when $\hat\kappa_i^c\approx \kappa_i^c$. 
When the candidate critical values were further changed and entered the actual 
post-crisis regime, we observed concave upward curves (blue curves with circles in panels 
(a1) and (a2)).

We built upon these observations to predict the actual crises values---having 
observed only the dynamics in the pre-crisis regimes---as follows. For ranges of 
$\kappa_i$ of the pre-crises regimes of both systems, we determined the 
logarithmic growth rates $g_i(\kappa_i)$ (see Eq.~\eqref{eq:logampl}). We then 
consider intervals of candidate critical values, $\hat{\kappa}_1^c$ and 
$\hat{\kappa}_2^c$, which are shown as abscissae in panels (b1) and (b2), 
respectively. For each candidate critical value $\hat{\kappa}_i^c$ of the 
interval, we consider the growth rate $g_i$ as a function of the logarithmic 
distance between $\hat{\kappa}_i^c$ and $\kappa_i$ and fitted a line. The mean 
squared error $\xi_i$ of this fit and its dependence on the candidate critical 
value is shown in (b1) and (b2). Both panels complement the previous 
observations: We observed the mean squared errors to become minimal (within the 
investigated intervals) when the candidate critical values equaled the actual 
critical values (marked as red circles in both panels). The linear scaling 
behavior revealed the actual critical values.

The accuracy of the prediction of the critical value will depend on various 
factors such as the set of control parameters for which the system dynamics was 
observed (from now on called ``observed parameters''). In particular, we 
hypothesized our predictions to become more accurate the closer the observed 
parameters are to the actual critical value. To study this hypothesis, we 
created a set $J$ of observed parameters, consisting of 50 equidistantly spaced 
control parameter values in the pre-crisis regime. With $\delta_i$ we denote the 
minimum distance between the actual critical value for System $i$ and any of the 
observed parameters of the set. Furthermore, we created a set of candidate 
critical values. For each of the candidate critical values, we considered the 
growth rates $g_i$ as a function of the logarithmic distance between the 
candidate critical value and $\kappa_i\in J$ and fitted a line. We predicted the 
critical value by determining the candidate critical value $\hat\kappa_i^c$ for 
which the mean squared error $\xi$ of this fit was the smallest. Let 
$\hat{\kappa}_i^p$ denote this predicted value. We repeated this procedure for 
different sets $J$ which we created by adding a small value 
($\delta\tilde{\kappa}_i$) to all members of the previous set, thereby yielding 
sets of observed parameters with different distances $\delta_i$ to the actual 
critical value. 

In figure~\ref{fig:2} (c1) and (c2), we show the dependence of the predicted 
critical values on the distance of the set of observed parameters. Indeed, for 
small distances $\delta_i$, we observed the predicted values to be very close to 
the actual critical values. Increasing $\delta_i$ and thus observing the 
dynamics for control parameters that are more distant from the actual critical 
values ($\delta_1 > 10^{-4}$, $\delta_2 > 3.5\cdot 10^{-6}$), we observed our 
predicted critical values to increasingly deviate from the actual ones, tending 
towards the pre-crisis regime. Consequently, in our numerical experiments, the 
probability to falsely locate the transition in the actual post-crisis regime 
(and thus risking to enter such a regime without warning) is marginal. These 
results underline that the predictions of an interior crisis become more 
accurate the closer we observe our dynamics with respect to the transition, an 
observation also reported with respect to local bifurcations and abrupt monsoon 
transitions~\cite{Thomas2015}.

\begin{figure}[t]
\includegraphics[width=0.99\linewidth]{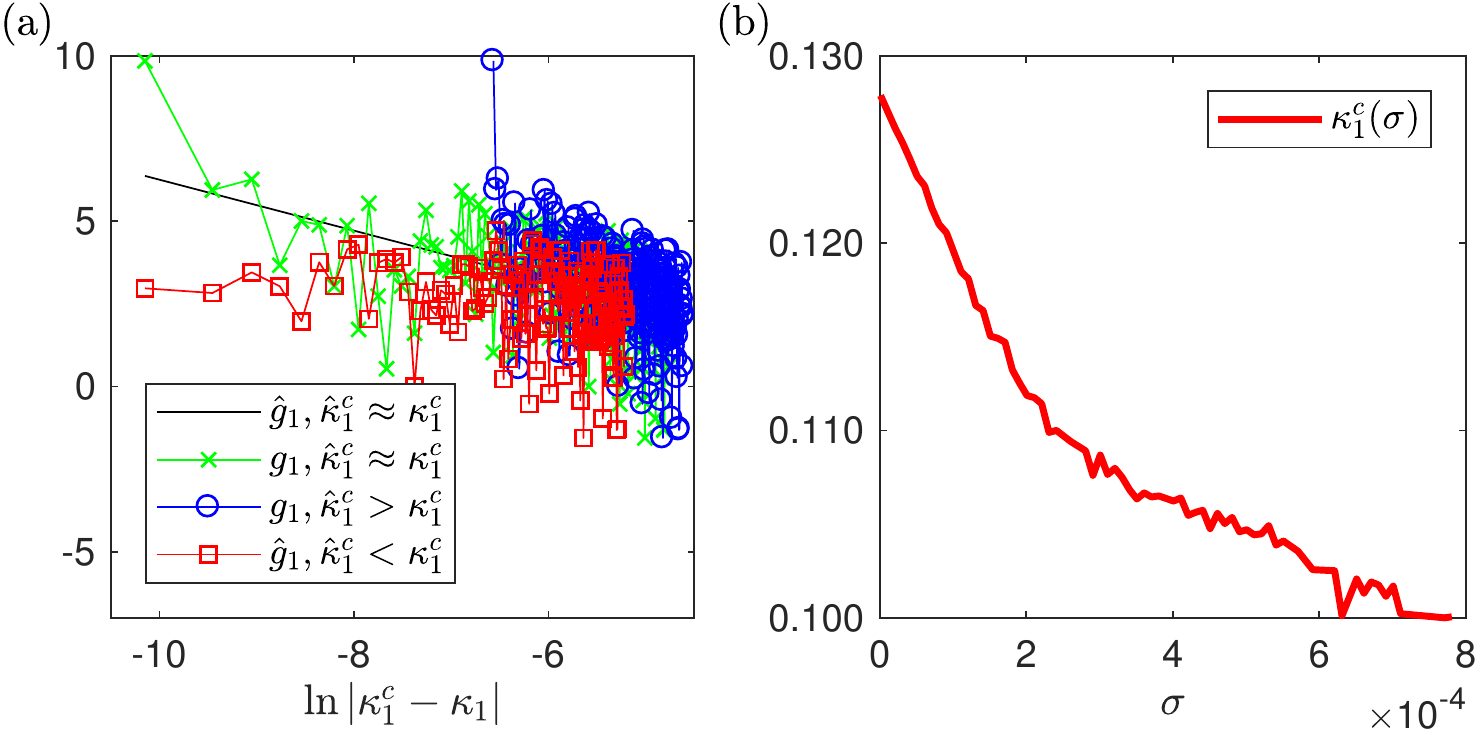}%
\caption{(a) Logarithmic growth rate $g_{1^\prime}$ for values 
$\kappa_{1^\prime}$ of the pre-crisis regime of System~1' with noise level 
$\sigma=10^{-4}$. $g_{1^\prime}$ scales approximately linearly with the 
logarithmic parametric distance between the actual crisis value 
$\kappa_{1^\prime}$ and candidate critical values $\kappa_{1^\prime}^c$ (green crossed 
line: $\kappa_{1^\prime}^c \approx 0.118593$; black continuous line: linear fit with slope 
$-0.771$). Candidate critical values below (above) the actual critical yield 
concave downward (upward) curves with $\hat{\kappa}_{1^\prime}^c \approx 
0.115469$ ($\hat{\kappa}_{1^\prime}^c \approx 0.119961$). System~1' was 
numerically solved using the Euler-Maruyama method (step size $\delta t=0.001$). 
For each values $\kappa_{1^\prime}\in [0.11,0.12]$ (step size 
$\delta\kappa_{1^\prime} \approx 3.90625\times 10^{-5}$), an ensemble of 
$5\times10^3$ initial conditions was created and evolved for $4\times10^8$ time 
units. The initial $10^8$ time units were evolved without noise and discarded as 
transients. From the ensemble, the global maximum amplitude (of the mean value 
$\bar{x}^\prime$ of both excitatory variables of System~1') which enter 
equation~\eqref{eq:logampl} was determined. (b) The critical value 
$\kappa_{1^\prime}^c$ decreases with increasing noise level $\sigma$. To 
determine $\kappa_{1^\prime}^c$ as a function of $\sigma$, a similar simulation 
protocol as for (a) was used but with an ensemble of $10^3$ initial conditions 
for each $\kappa_{1^\prime} \in [0.1,0.128]$ ($\delta\kappa_{1^\prime} \approx 
5.46875\times 10^{-5}$) and $\sigma\in[10^{-6},10^{-2}]$ ($\delta \sigma = 
10^{-5}$). For each initial condition, the system was evolved for $10^8$ time 
units, and initial $10^7$ time units were discarded as transients. For a given 
value of $\sigma$, the smallest value of $\kappa_{1^\prime}$ for which we 
observed $\bar{x}^\prime>0.5$ was picked as the actual critical value 
$\kappa^c_{1^\prime}$.
\label{fig:3}}
\end{figure}

\begin{figure}
\includegraphics[width=\linewidth]{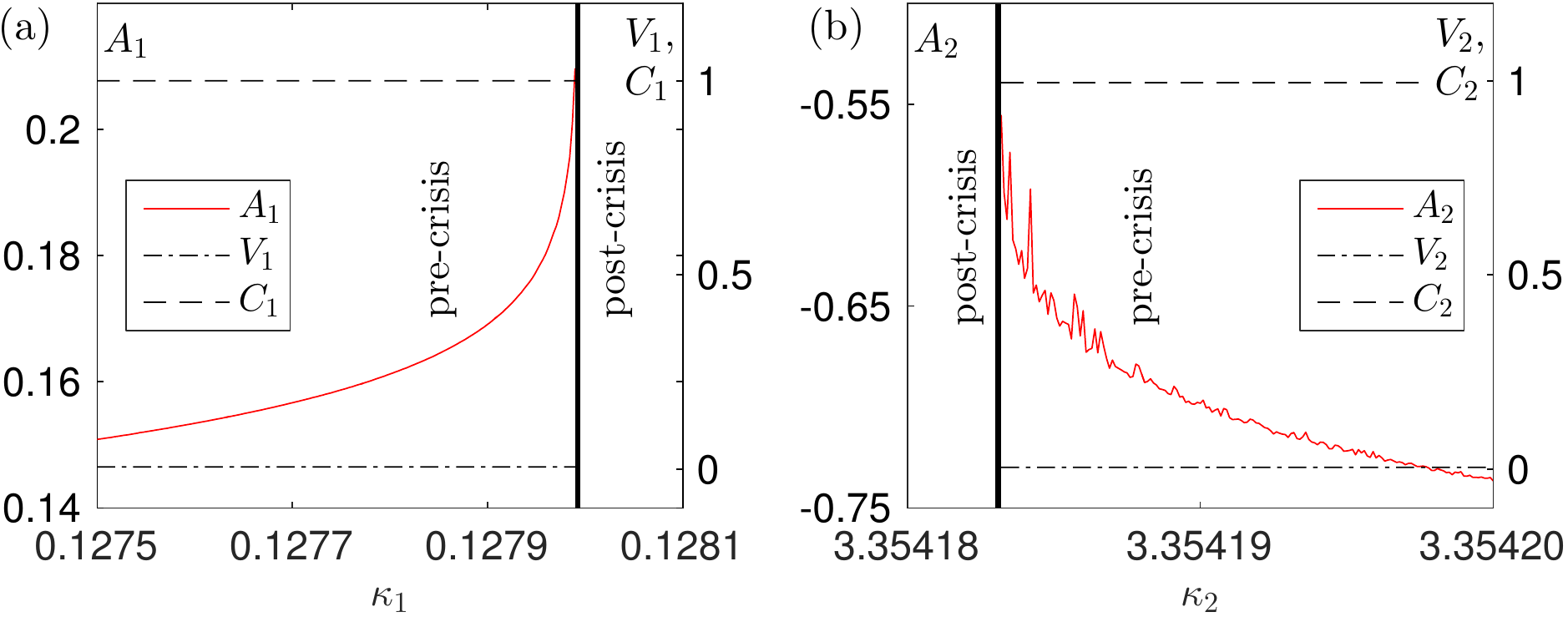}%
\caption{Maximum amplitudes $A_1$ and $A_2$ (red continuous lines) for System~1 (panel (a)) 
and System~2 (panel (b)) as functions of the control parameters $\kappa_1$ and 
$\kappa_2$, respectively. Variances $V_i$ and lag-1 autocorrelations $C_i$ are 
shown as dash-dotted and dashed black lines, respectively. For each value of 
$\kappa_i$, $V_i$ and $C_i$ were determined from a single time series of 
$\bar{x}$ (System 1) and $u$ (System 2), e.g., for System~2, $V_2 = 
T_2^{-1}\sum_{t=1}^{T_2}\left( u(t)-\bar{u}\right)^2$, 
$C_2=V_2^{-1}T_2^{-1}\sum_{t=1}^{T_2-1}\left(u(t)-\bar{u}
\right)\left(u(t+1)-\bar{u}\right)$. $V_1$ and $C_1$ are defined analogously 
($T_1=2\times10^4$, $T_2=10^5$). In each panel, the left ordinate specifies the 
scale of the maximum amplitudes whereas the right ordinate specifies the scale 
of the variance and the lag-1 autocorrelation. Black vertical lines mark the 
locations of the critical values.
\label{fig:4}}
\end{figure}

The dynamics we considered so far were purely deterministic, a property that is 
not ubiquitously present in systems investigated in the field. In field studies, 
measurements may be contaminated by noise contributions and the investigated 
dynamics itself may contain stochastic components (for instance, consider 
biological systems and the intrinsic stochastic nature of involved processes). 
While increasing levels of measurement noise will likely decrease the parameter 
regime before the crisis for which we can reliable predict the critical 
transition (due to noise masking the maximum amplitudes), the effect of 
stochastic components present in the system dynamics on our prediction abilities 
could be non-trivial. To study this question, we added a Gaussian white noise 
term $\eta$ to the first rate equation of System~1 and refer to this ``noisy'' 
system as System~1', $\dot{x}_1^\prime = x_1^\prime (a-x_1^\prime) 
(x_1^\prime-1) - y_1 + \kappa_1 (x_2 - x_1^\prime)+\sigma\eta(t)$, where 
$\sigma^2$ is the noise strength. Like for Systems~1 and 2, we observed a linear 
scaling behavior of the growth rate with respect to the logarithmic parametric 
distance (cf. figure~\ref{fig:3}(a), $\sigma=10^{-4}$) when the candidate 
critical value was equal or arbitrarily close to the actual critical value. This 
scaling behavior allowed us to predict the critical transition even in 
System~1', a result highlighting the robustness of our reported scaling 
characteristic. 

The actual critical value in System~1' is smaller than the critical value in 
System~1 (for $\sigma>0$) and decreases with increasing noise strength 
(figure~\ref{fig:3}(b)), thereby allowing System~1' to earlier enter the 
post-crisis regime. The post-crisis dynamics showed rare and recurrent high 
amplitude oscillations like those reported for System~1 in previous 
studies~\cite{Ansmann2013,Karnatak2014b}. These events are mediated by a 
channel-like structure in state space, through which a trajectory can escape for 
a long excursion associated with an extreme event. Our results highlight that 
noise in the pre-crisis regime can facilitate the access of the trajectories to 
this channel-like structure, thereby in a sense, lowering the critical value. 
Similar observations were also reported in optical systems where noise induced 
rogue waves were observed in parameter ranges which were pre-crisis for the 
equivalent system without noise~\cite{Zamora-Munt2013}.

Finally, we demonstrate that the predictive information about the impending 
global bifurcations we considered here is contained in the maximum amplitudes of 
signals but not in their variance or lag-1 autocorrelations. The latter two are 
early warning signals that have been frequently considered in the framework of 
critical slowing down and are expected to increase before a local bifurcation. 
In figure~\ref{fig:4}~(a) and (b), we show the maximum amplitudes $A_1$ and 
$A_2$ for Systems~1 and 2, respectively, prior to the interior crises. In 
addition, we overlayed these graphs with the variances $V_i$ and lag-1 
autorcorrelations $C_i$ obtained from observables $\bar{x}$ (System 1) and $u$ 
(System 2). Approaching the crises, we did not observe $V_i$ and $C_i$ to change 
drastically and, hence, to indicate any impending transition. We are also not 
aware of any theoretical argument regarding why both should show changes prior 
to a crisis. We observed both quantities to stay approximately constant. In 
contrast, the maximum amplitudes $A_i$ increased prior to the crises---a 
phenomenon we call \emph{critical attractor growth}---, indicating the impending 
interior crises. 

Why do we observe critical attractor growth in these excitable systems? We 
hypothesize that the geometric properties of such systems are the primary reason 
behind the observed growth in amplitudes. The changes in the alignment of the 
invariant manifolds with parameter variation (as discussed in previous 
studies~\cite{Ansmann2013,Karnatak2014b}) and the related changes in the flow 
field patterns allow the trajectories to evolve into the parts of state space 
which they could not access for parameter values more distant to the crisis, 
thereby exhibiting the observed growth in maximum amplitudes. We believe that 
future work that builds upon our findings and provides mathematical results 
highlighting the mechanisms underlying critical attractor growth has a strong 
potential to reveal how early warning signals for global bifurcations and other 
classes of dynamical systems can be derived, thereby increasing the set of 
critical transitions that can be predicted.

In summary, we introduced an early warning signal that is sensitive to an 
upcoming global bifurcation, namely an interior crisis in excitable systems. 
This extends the existing toolbox of early warning signals beyond the prediction 
of local bifurcations, a need also put forward in previous 
works~\cite{Scheffer2012,Boettiger2013}. We expect two directions of research to 
be particularly fruitful: Achieving a theoretical understanding of our early 
warning signal might pave the way towards the development of signals that are 
sensitive to even a larger set of global bifurcations. Moreover, for systems 
moving gradually towards a transition, the development of frameworks that allow 
to reliably translate early warning signals from monitoring time series into the 
predictions of points in time at which a transition will occur would be very 
useful. They would likely open up early warning signals to a plethora of 
applications, enabling us to be better prepared and to make better informed 
decisions with respect to impending critical transitions in a variety of 
dynamical systems that are relevant to human societies.

\begin{acknowledgments}
S.B. acknowledges support by the Volkswagen foundation (Grant No. 88461).
\end{acknowledgments}

\end{document}